\documentclass[aps,prl,twocolumn,groupedaddress,showpacs,amsmath,amssymb]{revtex4}
\newcommand\thintilde{{\lower.74ex\hbox{\mathtt{\char`\~}}}}
\usepackage{graphics}
\usepackage{color}
\usepackage{graphicx}
\usepackage{natbib}
\newcommand{\un}[1]{\ensuremath{\, \mathrm{#1}}}
\begin{document}

\title{Synchronization of Coupled Boolean Phase Oscillators}

\author{David P. Rosin,$^{1,2}$ Damien Rontani,$^{1,3}$ and Daniel J. Gauthier$^1$}

\affiliation{$^1$Department of Physics, Duke University, 120 Science Drive, Durham NC~27708, USA\\
$^2$Institut f\"ur Theoretische Physik, Technische Universit\"at Berlin - Hardenbergstr.~36, Berlin D-10623, Germany\\
$^3$Sup\'{e}lec, OPTEL research group and LMOPS EA-4423, 2 Rue Edouard Belin, Metz F-57070, France}
\thanks{}
\date{\today}
\begin{abstract}
We design, characterize, and couple Boolean phase oscillators that include state-dependent feedback delay. The state-dependent delay allows us to realize an adjustable coupling strength, even though only Boolean signals are exchanged. Specifically, increasing the coupling strength via the range of state-dependent delay leads to larger locking ranges in uni- and bi-directional coupling of oscillators in both experiment and numerical simulation with a piecewise switching model. In the unidirectional coupling scheme, we unveil asymmetric triangular-shaped locking regions (Arnold tongues) that appear at multiples of the natural frequency of the oscillators. This extends observations of a single locking region reported in previous studies. In the bidirectional coupling scheme, we map out a symmetric locking region in the parameter space of frequency detuning and coupling strength. Because of large scalability of our setup, our observations constitute a first step towards realizing large-scale networks of coupled oscillators to address fundamental questions on the dynamical properties of networks in a new experimental setting.
\end{abstract}



\maketitle

\section{I. Introduction} Collective behaviors in networks of coupled systems have been thoroughly studied experimentally using mechanical, electronic, and optoelectronic setups \cite{ROY10,MAR13,WIL13,NIX11}. Here, we describe a new dynamical system that is based on autonomous logic circuits that can potentially be assembled into a large-scale network for the study of collective dynamics. Dynamical behaviors in networks, such as synchronization, depend critically on the strength of coupling of the nodes and hence the availability of an adjustable coupling strength is key. However, defining a coupling strength in the framework of autonomous Boolean systems is difficult as they allow for the exchange of signals with only two Boolean levels  (for example ``ON''/``OFF'' or ``1''/``0''). Therefore, the coupling is restricted to all-or-nothing in most Boolean networks, such as that described in our previous work \cite{ROS13c}.  As a solution, we demonstrate here that an effective tunable coupling strength can be implemented in an autonomous Boolean system based on state-dependent delay.

Autonomous logic circuits have already been used to build experimental networks of excitable and chaotic systems \cite{ROS13a,ROS13b,ROS13c,ZHA09}. These networks---an experimental realization of autonomous Boolean networks---evolve continuously in time and display dynamics that are governed by the network topology, the logic functions of each node, and the link time delays \cite{GLA96,GLA98}. The dynamical states of the nodes, on the other hand, take on Boolean values so that interactions specific to the amplitude of the dynamics are usually considered not important. However, in the experimental realization, non-ideal behaviors, such as short-pulse rejection, can play a crucial role in certain dynamical regimes \cite{ZHA09}.

We extend the framework of autonomous logic circuits by realizing dynamical nodes with oscillatory behavior and tunable frequency. Specifically, we describe the design of a Boolean phase oscillator that can switch between two free-running frequencies, similar to device designs coming out of previous research on phase-lock loops (PLLs) \cite{BES03,HSU99}. The oscillator is a Boolean feedback system with state-dependent feedback delay. We demonstrate experimentally and numerically that such oscillators can synchronize (phase-lock) in uni- and bidirectional coupling schemes. In both cases, we map out domains of synchronization in parameter space and demonstrate that the difference in free-running frequencies is analogous to a coupling strength encountered in traditional weakly coupled oscillators studied by the nonlinear dynamics community \cite{PIK01}.

\section{II. Designing a Boolean Phase Oscillator}

\subsection{A. Definition of phase oscillators and example of physical realizations} A phase oscillator is a mathematical construction resulting from the continuous one-to-one mapping of a limit cycle in phase space onto the unit circle \cite{PIK01}. The phase information of a square-wave oscillator can be extracted directly from its scalar periodic time series $V(t)$ by either calculating the phase at rising and falling transitions or by converting the time series into low-frequency Fourier modes using a low-pass filter. Therefore, we refer below to a periodic square-wave oscillator as a Boolean phase oscillator (BPO) where the voltage alternates between two Boolean values. One type of BPO that can be coupled to external signals using Boolean elements is an all-digital phase-locked loop (ADPLL). These systems are widely used in digital communication to achieve frequency multiplication and clock synchronization \cite{BES03}. Based on these widely established  PLLs, we design a BPO that includes state-dependent time delay.

\subsection{B. Boolean phase oscillators and phase-locked loops (PLLs)}
\begin{figure}[b!]
\begin{center}
\includegraphics[width=7cm]{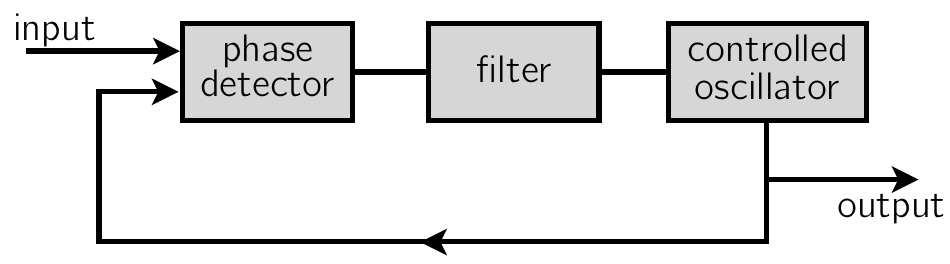}
\end{center}
\caption{\label{fig:PRE_Fig_2New}
Graphical description of the three principal functional blocks comprising a phase-locked loop (PLL). }
\end{figure}

PLLs comprise three functional blocks: (i) a phase-detector  block, (ii) a filter block, and (iii) a controlled-oscillator (CO) block. These three blocks are assembled as shown in Fig.~\ref{fig:PRE_Fig_2New}. The phase detector generates an error signal that is proportional to the phase difference between the output signal of the PLL and a reference signal. The error signal is filtered before being applied to the CO block, which adjusts its phase and frequency \cite{BES03}.

Recently, advances in digital electronic systems have spurred the development of all-digital PLLs (ADPLLs), where all three blocks shown in Fig.~\ref{fig:PRE_Fig_2New} are realized with electronic logic circuits \cite{ARA06}. The phase detector in ADPLLs typically generates $N$-bit numbers representing the phase shift between the input and output signals. The filter in ADPLLs is usually implemented with a digital up-down counter that integrates the signal.

While there has been much effort in improving the locking performance of PLLs, our goal here is to develop a very simple, resource-efficient PLL design that allows us to create large networks. In the following section, we detail how to realize such a simple circuit.

\subsection{C. Designing a simple Boolean phase oscillator}
\label{sec:design_of_BPO}
\begin{figure}[h!]
\begin{center}
\includegraphics[width=8cm]{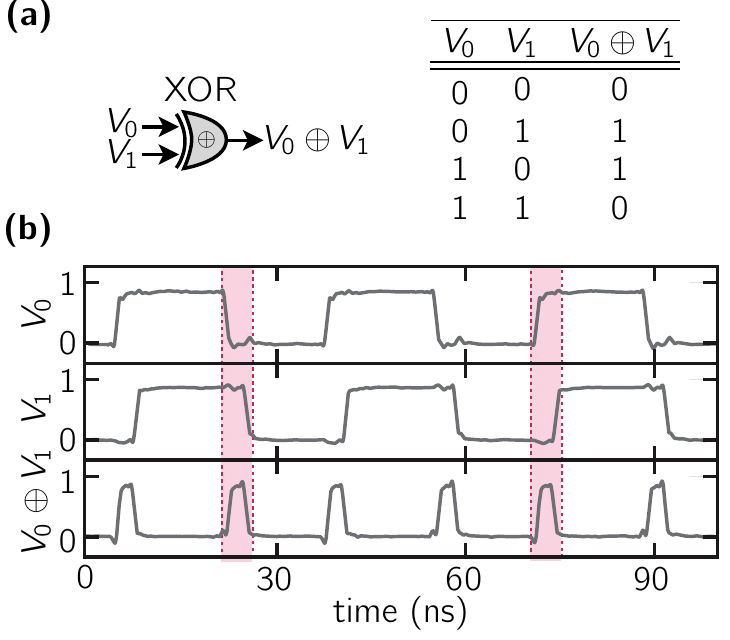}
\end{center}
\caption{\label{fig:parts_dynamics}
(Color online) (a) Symbol of an  XOR logic gate that is used as a one-bit phase detector and its associated look-up table. The low ($V_L$) and high ($V_H$) Boolean voltage are denoted by symbols $``0"$ and $``1"$ respectively. (b) The error signal $V_0\oplus V_1$ resulting from the phase detector with two phase shifted input signals $V_{0}$ and $V_{1}$ of frequency $f_0 = f_1 = 30.0 \pm 0.1 \text{ (MHz)}$. The phase difference is highlighted by shaded regions at a falling and a rising Boolean transition. The waveform is obtained from an experimental implementation on a field-programmable gate array (FPGA), specifically the model Altera Cyclone IV EP4CE115F29C7, which is used for all experimental implementations in this paper.}
\end{figure}

The design of our Boolean phase oscillator follows the structure of three functional blocks shown in Fig.~\ref{fig:PRE_Fig_2New}. For the phase detection block, we use a two-input XOR logic gate that is a single-bit digital phase detector [see Fig.~\ref{fig:parts_dynamics}(a)] \cite{BES03}. The waveform of the resulting error signal $V_c$ is shown in Fig.~\ref{fig:parts_dynamics}(b) for two Boolean input waveforms of different phase and equal frequency. The error signal satisfies $V_c=V_{H}$ (Boolean signal ``1") when the two input signals have different Boolean values and $V_c=V_{L}$ (Boolean signal ``0") otherwise. This results in a pulse-shaped waveform at the output of the phase detector with a pulse width proportional to the phase difference between the two inputs as highlighted in the figure. Note that the XOR function does not give the sign of the phase difference, only its absolute value.

The second block, the filter, is not explicitly implemented in our design. Nevertheless, each logic gate low-pass filters intrinsically the voltage generated by the phase-detection block with a cutoff frequency related to the gate propagation delay $\tau_{LG}=0.275\pm0.010\un{ns}$.

\begin{figure}[b!]
\begin{center}
\includegraphics[width=7.75cm]{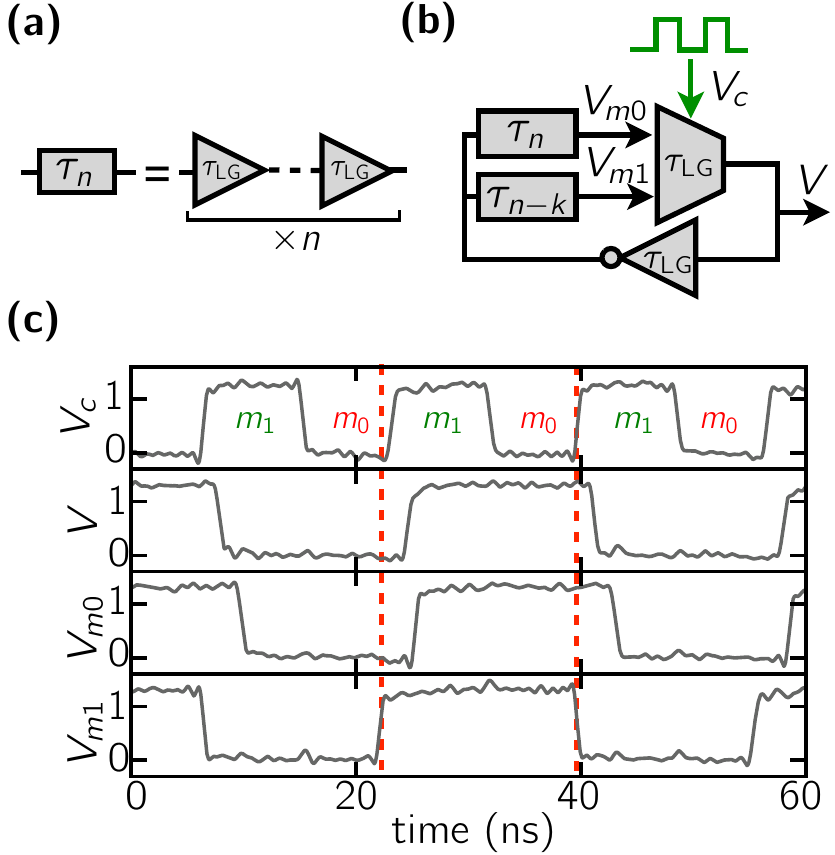}
\end{center}
\caption{\label{fig:PRE_Fig_4New}
(Color online) (a) Construction of delay line $\tau_n$ with a series of $n$ cascaded buffers with individual propagation time $\tau_\mathrm{LG}$. (b) Schematic of the CO block of our BPO. (c) Locking of the CO block to an externally-generated control signal $V_c$ with frequency $f=60.0\pm 0.1\un{MHz}$. Also shown are the output waveform $V$ and the signals from the delay lines that are input to the switch $V_{m0}$ and $V_{m1}$. The parameters of the CO block are $n=65$ and $k=10$. The gate propagation delay as measured for buffer gates is $\tau_{LG}=0.275\pm0.010\un{ns}$}
\end{figure}

Finally, the CO block is based on a simplified design proposed in Ref.~\cite{HSU99} and consists of an inverter gate with a state-dependent feedback delay. Specifically, it is realized using one inverter gate, two series of $n\in\mathbb{N}$ and $n-k\in\mathbb{N}$ cascaded buffer gates, and a Boolean switch as shown in Fig.~\ref{fig:PRE_Fig_4New}(a) and (b). In the resulting delayed feedback system with negative gain, the dynamics has a periodicity equal to twice the feedback delay because it requires two inversions to recover the initial Boolean state \cite{KAT98}. Therefore, the frequency of oscillation is
\begin{equation}
f(V_c)=\frac{1}{2\tau(V_c)},
\end{equation}
where $\tau(V_c)$ is the state-dependent feedback delay that depends on the control signal $V_c$. The cascaded buffer gates implement two feedback delays $\tau_{n-k}=(n-k)\tau_\mathrm{LG}$ and $\tau_k=k\tau_\mathrm{LG}$. The feedback lines can also be realized with cascaded inverter gates, but our choice of buffer gates results in enhanced low-pass filtering (see Section III.B for a detailed explanation). The switch, implemented as a three-input logic gate with the look-up table of a multiplexer, selects between the two feedback delays $\tau_{n-k}$ and $\tau_{n}=\tau_{n-k}+\tau_{k}$ depending on the control signal $V_c$, which results in a state-dependent feedback delay of
\begin{equation}
\tau (V_c) = \left\{ {\begin{array}{*{20}c}
   {\tau_n} &{=}&{n \tau_\text{LG}} & {\text{if } V_c\le V_{th},}  \\
   {\tau_{n-k}}&{=}&{(n- k)\tau_\text{LG}} & {\text{otherwise}.}  \\
 \end{array} } \right.
\end{equation}
Here, the integer $k\approx(\tau_n-\tau_{n-k})/\tau_\mathrm{LG}$ is proportional to the difference of the two possible values of the feedback delay. When a constant control voltage of $V_c = V_{L}$ or $V_c = V_{H}$ is applied to the CO, then it will oscillate at a free-running frequency of $f_n = 1/2\tau_n$ or $f_{n-k} = 1/2\tau_{n-k}$, respectively.

The dynamics of the CO block is characterized by sending an external signal of frequency $f_c=60.0\pm0.1\un{MHz}$ to the $V_c$ port. Figure.~\ref{fig:PRE_Fig_4New}(c) shows the external waveform, the resulting output voltage of the CO block $V(t)$, and the output voltages of the two delay lines $V_{m0}(t)$ and $V_{m1}(t)$. These two signals are time shifted by the delay $\tau_k$ so that $V_{m1}(t) = V_{m0}(t+\tau_k)$. In the figure, the output voltage is $V(t)=V_\mathrm{m0}(t)$ [$V(t) = V_\mathrm{m1}(t)$] when $V_c=V_\mathrm{L}$ [$V_c = V_\mathrm{H}$], which is due to the functionality of the switch in the setup. As a result, a rising-edge Boolean transition in $V_c$ [highlighted with dashed lines in Fig.~\ref{fig:PRE_Fig_4New}(c)] leads to a positive phase shift, which triggers rising and falling edges in the output voltage $V(t)$. Therefore, switching between the two feedback delays leads to frequency locking. For more details, see Ref.~\cite{HSU99}.

\section{III. Synchronization of Boolean Phase Oscillators}

\subsection{A. Unidirectional synchronization of experimental Boolean phase oscillators and weak coupling analogy}
By combining the phase detector from Fig.~\ref{fig:parts_dynamics}(a) with the CO block from Fig.~\ref{fig:PRE_Fig_4New}(b), we obtain the Boolean phase oscillator (BPO). We use the BPO in the rest of the paper to study synchronization of coupled oscillators.

\begin{figure}[b]
\begin{center}
\includegraphics[width=8.5cm]{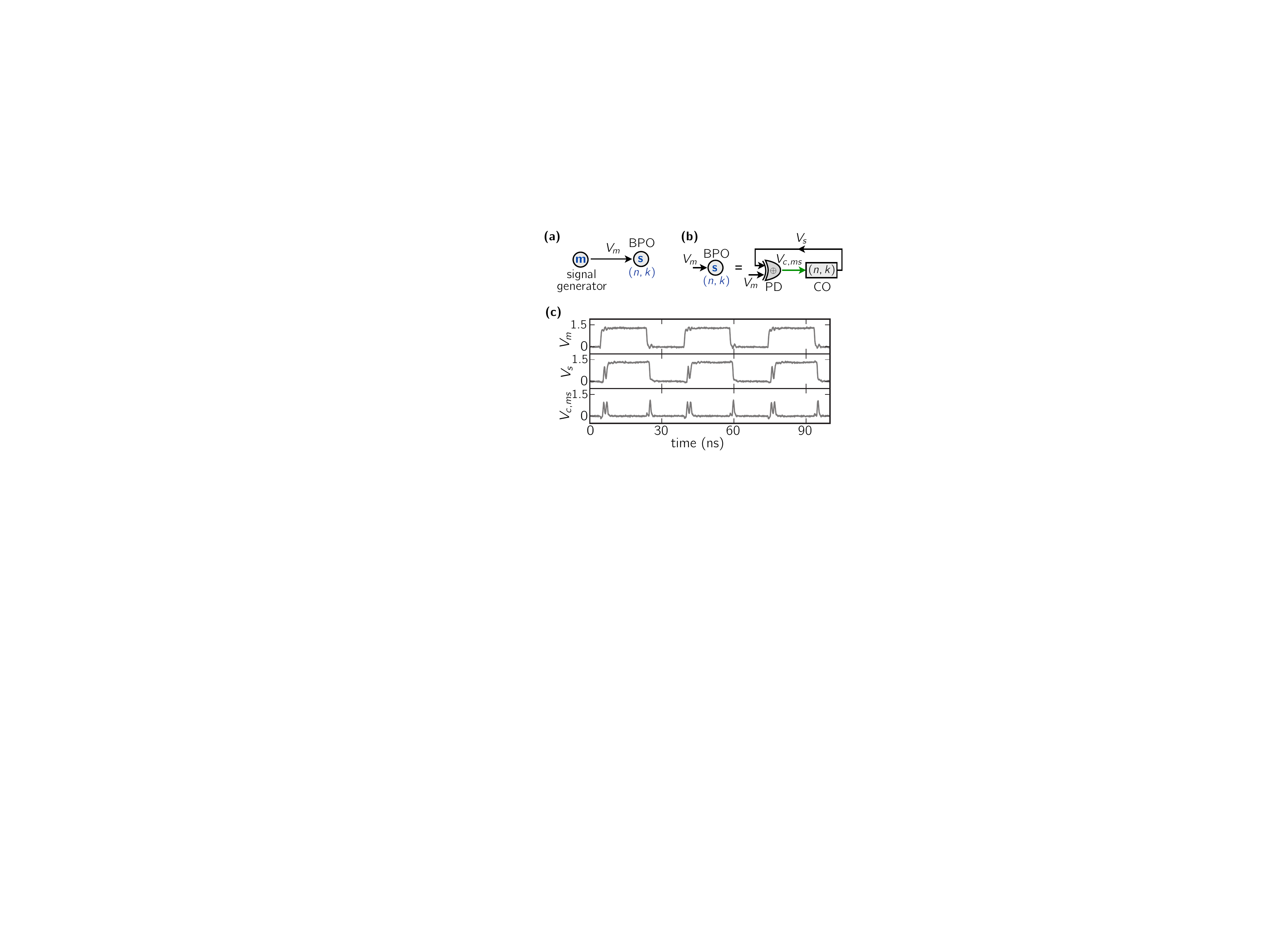}
\end{center}
\caption{\label{fig:master_slave}
(Color online) (a) Experimental setup of the master-slave coupling scheme for two BPOs. (b) Construction of the slave BPO comprising the CO block labeled $(n,k)$ [see Fig.~\ref{fig:PRE_Fig_4New}] and an XOR-based phase detector (PD) block. (c) Waveforms of the master oscillator (generated externally with waveform generator Tektronix AFG3251) $V_{m}$ of frequency $f_m=28.6\pm0.1\un{MHz}$, the output waveform of the slave oscillator $V_{s}$, and the error signal $V_{c,ms}=V_m\oplus V_s$. The parameters of the CO block in the BPO are the same as the ones in Fig.~\ref{fig:PRE_Fig_4New}.}
\end{figure}

We test the locking capabilities of the BPO with an external driving signal of frequencies $f_m$. This setup, shown in Fig.~\ref{fig:master_slave}(a)-(b), can be interpreted as a master-slave configuration, where two BPOs are coupled unidirectionally. Here, the equivalent master BPO is an external function generator, which provides finer frequency control of $f_m$ than changing discretely (by adding or removing buffers in the ring) the frequency of the master BPO.

In Fig.~\ref{fig:master_slave}(c), we show the dynamics resulting from the master-slave configuration. Specifically, we show the phase-locked waveforms of the master and slave oscillator with frequencies $f_m = f_s =28.6 \pm 0.1\text{ MHz}$, and the error signal $V_c$. Here, the dynamics of the BPO leads to a constant phase shift between master and slave oscillator that results in a pulsed signal $V_c$ [similar to Fig.~\ref{fig:parts_dynamics}(b)]. The pulses in $V_c$ provide the phase correction that allows synchronization. The waveforms of $V_s$ and $V_c$ display fast oscillations at the Boolean transitions, which are due to unfiltered feedback between the phase detector and the control port of the CO block.

\begin{figure}[tb]
\includegraphics[width=8.5cm]{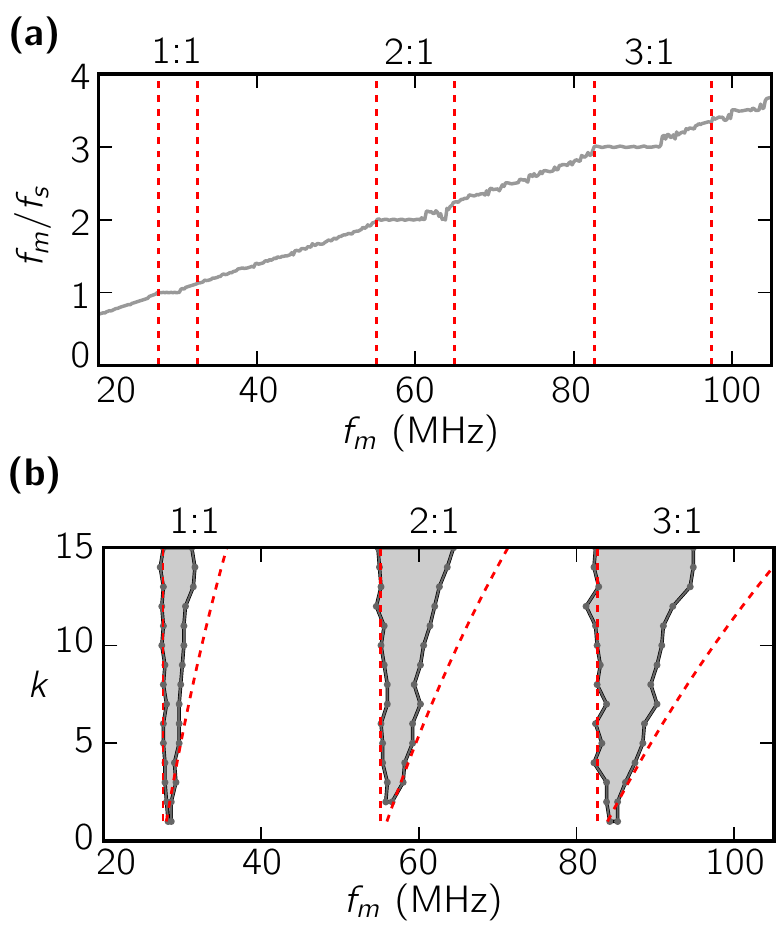}
\caption{\label{fig:03}
(color online) (a)~Experimental Devil's staircase for a BPO with $k=10$ showing synchronization regions as horizontal lines. For the three integer synchronization regions $f_\mathrm{m}{:}f_\mathrm{s}=1{:}1$, $2{:}1$, and $3{:}1$, the theoretical limits are shown with red dashed lines. (b)~Experimental Arnold tongues for the three integer synchronization regions as a function of $k$ and the theoretical limits of the regions marked with red dashed lines. Other Arnold tongues are not shown. Setup and experimental parameters as in Fig.~\ref{fig:PRE_Fig_4New}.}
\end{figure}

We repeat the coupling experiment by tuning the frequency of the master oscillator $f_m$ from $20$ to $105\un{MHz}$ while keeping the parameters of the slave BPO unchanged. We measure $f_s$ and calculate the ratio $f_\mathrm{m}/f_\mathrm{s}$ as a function of $f_\mathrm{m}$. This measurement leads to the so-called \emph{devil's staircase} shown in Fig.~\ref{fig:03}(a) \cite{NOR99,PIK01}. In the graph, synchronization regions with constant ratios $f_\mathrm{m}/f_\mathrm{s}$ are represented by horizontal plateaus---the \textit{stairs}. The most prominent synchronization regions are associated with integer ratios $f_\mathrm{m}{:}f_\mathrm{s}=p{:}q$ with $q=1$ and $p=1,2,3$. In the synchronization region $2{:}1$, $f_\mathrm{m}/f_\mathrm{s}\neq2$ for a narrow range of $f_m$. This imperfection appears also in numerical simulations as discussed below. Moreover, many narrow fractional synchronization regions $p{:}q$ exist.

The widths of the synchronization regions---the stairs---are known as the locking range $U$ \cite{HSU99}. The maximum theoretical locking range $U_{max}$ can be determined, so that $U\subseteq U_{max}$ with $U_{max}=\left[f_n,f_{n-k}\right]$ for the first integer synchronization region $1{:}1$ with free-running frequencies of the BPO $f_n$ and $f_{n-k}$. Then, the locking range is given by $f_m\in U_1\subseteq\left[f_n,f_{n-k}\right]$. For integer synchronization regions of higher order $p{:}1$, the frequency locking of master and slave is given by $f_m=pf_s$. This leads to larger synchronization regions of $U_p\subseteq\left[pf_n,pf_{n-k}\right]$. In Fig.~\ref{fig:03}(a), we display the theoretical boundaries $\left[pf_n,pf_{n-k}\right]$ with dashed lines. The figure shows that the right theoretical locking boundaries are not reached in the experiment. 

The non-saturation of the right theoretical boundaries in the $p{:}1$ synchronization region originates from the generation of the control signal $V_{c,ms}$ with little frequency filtering. Specifically, the right boundary corresponds to large detuning between the frequency of the master oscillator $f_m$ and and the free-running frequency of the slave oscillator $f_n$. Then, the control signal $V_{c,ms}$ displays the Boolean voltage $V_{H}$ for an increased time to allow for greater frequency adjustment. However, this also results in increased high-frequency oscillations due to unfiltered feedback in the oscillator as discussed above and visible in Fig.~\ref{fig:master_slave}(c). These high-frequency oscillations decrease the locking abilities when the frequency detuning is large.

To complete our analysis of the locking range, we also measure them as a function of $k$. In Fig.~\ref{fig:03}(b), we map out the integer synchronization regions and their analytic boundaries as a function of $k$ and $f_m$. Here, the system parameter $k$ is proportional to the variability in state-dependent delay $\tau_k=k\tau_\mathrm{LG}$ in the CO block. The resulting synchronization regions, also known as \emph{Arnold tongues} \cite{NOR99}, have a triangular shape that opens with increasing $k$. The analytic boundaries corresponding to $pf_n$ are shown with dashed lines. The Arnold tongues are subject to experimental variation of $\pm 3.5\%$ due to variations of gate propagation delays $\tau_\mathrm{LG}$ that appear in the experiment when implementing the oscillators on different location on the electronic chip called field-programmable gate array (FPGA) for different values of $k$.

The increase of synchronization regions $U_p$ with $k$ allows to draw an analogy of $k$ to the coupling strength of phase oscillators because they display similar synchronization regions that increase with the coupling strength \cite{PIK01}. Therefore, we have shown that adjustable weak coupling is possible in our BPO similar to the Kuramoto model \cite{KUR84}.

The mechanism to generate the coupling is via state-dependent delay as described in Section~II.C. Specifically, the phase detector generates a high Boolean voltage $V_c(t) = V_H$ with a duration proportional to the phase difference between the input periodic signal---generated either by another oscillator or a signal generator---and the output signal of the BPO. The signal $V_c(t)$ induces an increase in frequency of the BPO because the feedback delay switches to a shorter value. The magnitude of the resulting frequency adjustment is proportional to $k$. Therefore, larger values of $k$ lead to a larger frequency adjustment, which reduces the phase difference between the state of the BPO and the external periodic signal. This is analogous to the phase-correction mechanism encountered in theoretical models of phase oscillators \cite{PIK01}.

\subsection{B. Glass model for the Boolean phase oscillator}
We reproduce quantitatively our experimental results with a phenomenological model given by a set of piecewise-linear delay differential equations. This is an extension to an approach by Glass \textit{et~al.}~\cite{GLA97,GLA98} for the description of genetic networks in the framework of autonomous Boolean networks.

To model the master-slave setup, we describe the master oscillator simply by a periodic continuous square wave $y_m\in\left[-1,1\right]$ with low and high Boolean values $-1$ and $1$, respectively. These Boolean values can be scaled to correspond to the experimental Boolean voltages of $V_L=0\un{V}$ and $V_H=1.3\un{V}$. A BPO is described with continuous variables and associated Boolean variables which are both related via the threshold condition
\begin{equation}
  X(t)= -1 \text{ if } x(t) < x_\mathrm{th}; \text{ otherwise } X(t)=1,
\end{equation}
with $x(t)$ a continuous variable, $X(t)$ its Boolean counterpart, and $x_{th}=0$ the Boolean threshold value for $x(t)$.

In the original description by Glass \textit{et al.} \cite{GLA97}, each logic gate is modeled by a piecewise-linear differential equation. For the two delay lines in our setup that are based on $n$ buffers, this would result in $n$ differential equations only to represent the time delays. However, by instead including a state-dependent delay that accounts for the two delay lines in the CO, we can reduce the model of the BPO to two delay differential equations given by
\begin{subequations}
\begin{align}\label{eq:DE_main}
\tau_{x_s}\dot{x}_{s}  &= -x_s + [\lnot X_s(t-\tau_s(t))],\\
\label{eq:DE_filt}
\tau_{y_s}\dot{y}_{s} &= -y_s + [\lnot X_s(t-\tau_s(t))],
\end{align}
\end{subequations}
where $(x_s,X_s)$ are continuous and Boolean internal variables associated with the time delay lines of the slave BPO, $(y_s,Y_s)$ are the variable associated with the output of the slave BPO, $\lnot X=-X$ denotes the Boolean inversion (NOT) operation, $\tau_{x_s},\tau_{y_s}$ are the characteristic timescales associated with first-order low-pass filtering, and $\tau_s(t)$ is a state-dependent delay that we discuss in detail below.

Equation~\eqref{eq:DE_main} represents the low-pass filtering effect that results from the construction of the two delay lines with cascaded buffer gates. The characteristic timescale $\tau_{x_s}$ associated with the filtering is different for rising and falling transitions when measured for the CMOS-based logic gates \cite{WES00}. To account for this behavior, we include the following state-dependent switching condition depending on rise and fall
\begin{equation}\label{Eq:asymFilt}
\tau _{x_{s}}  = \left\{ \begin{array}{*{20}l}
  {\tau _\text{LG}/\ln 2}  &  {\text{if } x_{s}(t-\tau_{s}(t)) >   0,} \hfill \\
  {(\tau_\text{LG}/\ln2) \left(1 + n \Delta \tau_{rf}/\tau_\text{LG} \right)} &  {\text{if } x_{s}(t-\tau_{s}(t)) \le 0,} \hfill \\
\end{array}  \right.
\end{equation}
with $\Delta\tau_{rf}=24\pm2\un{ps}$ the time difference between rising and falling transitions of a single logic gate. The numeric value for $\Delta\tau_{rf}$ is measured by propagating a periodic square pulse through $n$ cascaded buffer gates that constitute a delay line. The measured output signal is changed according to Eq.~\eqref{Eq:asymFilt} resulting in a alteration of time difference between falling and rising transitions by $\tau_\text{LG}/(\ln2) (n \Delta \tau_{rf}/\tau_\text{LG})$. The accumulation of $\Delta\tau_{rf}$ produced by each logic gate results in pulse growth and enhanced low-pass filtering.

This non-ideal pulse-growth effect leads to filtering of high-frequency signals. Specifically, the Boolean-``1'' level of high-frequency modes extends until the whole waveform is Boolean-``1'' when propagating through the cascaded buffer gates. In contrast, such a filtering effect is reduced in delay lines based on cascaded inverter gates as their rise-fall asymmetry is inverted with every logic operation. This results in the appearance of higher harmonics in inverter-based ring oscillators \cite{LEE98}.

The dynamics of the switch in the CO is modeled with state-dependent delay according to the following condition 
\begin{equation}\label{Eq:switchdelay}
\tau _{s} (t) = \left\{ {\begin{array}{*{10}l}
   \tau_n & {\text{if } y_{c,ms}(t) \le 0,}  \\
   {\tau_{n-k}} & {\text{otherwise},}  \\
 \end{array} } \right.
\end{equation}
where $y_{c,ms}(t) = y_s(t - \tau _{c,ms} )\oplus y_m (t - \tau _{c,ms} )$. The additional delay $\tau_{c,ms}=2\tau_\mathrm{LG}$ accounts for the time that it takes for $V$ to propagate through the phase detector and reach the control port of the CO block (see also Fig.~\ref{fig:PRE_Fig_2New}). As a result, the delay switches between two values depending on the error signal $y_{c,ms}$ generated by the XOR-based phase detector [see Fig.~\ref{fig:parts_dynamics}(a) for the look-up table of the XOR operation]. 

In contrast to Eq.~\eqref{eq:DE_main}, Eq.~\eqref{eq:DE_filt} has no rising-falling asymmetry associated with the filtering as $\tau_{y_s}=\tau_\mathrm{LG}/\ln 2$ is constant. This is because its associated experimental voltage $V$ is measured after the switch and not after the delay lines that mainly contribute to the asymmetry. 

\begin{figure}[tb]
\includegraphics[width=8cm]{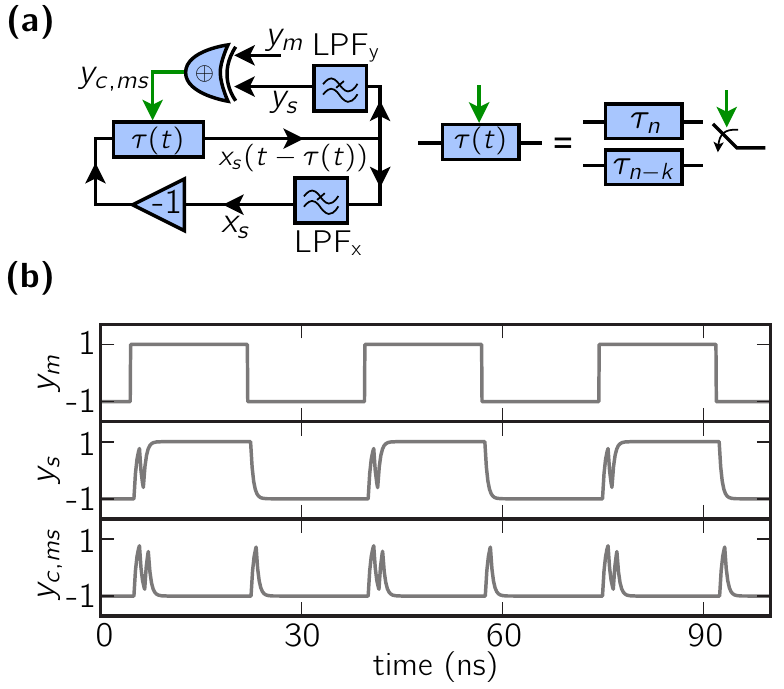}
\caption{
\label{fig:04}
(color online) (a) Block representation of the mathematical model for the master-slave setup. LPF$_{x,y}$  are low-pass filters with time constants $\tau_x$ and time constant $\tau_y$, respectively. (b) Numerical simulation with parameters as in Fig.~\ref{fig:PRE_Fig_4New}. The variables $y_m$, $y_s$, and $y_{c,ms}$ are the model representations of the experimental voltages $V_\mathrm{m}$, $V_\mathrm{s}$, and $V_\mathrm{c,ms}$ in Fig.~\ref{fig:master_slave}(b). The parameters are the same as in the experiment, see Fig.~\ref{fig:PRE_Fig_4New}. } 
\end{figure} 

Figure~\ref{fig:04}(a) is a graphical illustration of the model of the BPO in master-slave configuration. It is shown that the state-dependent delay of the slave oscillator is driven via the error signal $y_{c,ms}$, which is implicitly included in Eq.~\eqref{Eq:switchdelay} as an XOR operation of the signal $y_s$ and the output of the master oscillator $y_m$. The resulting variable $x_s$ is filtered by two different low-pass filters with and without asymmetric rise and fall times. Negative feedback results from the inverter gate. 

\begin{figure}[b]
\includegraphics[width=8.5cm]{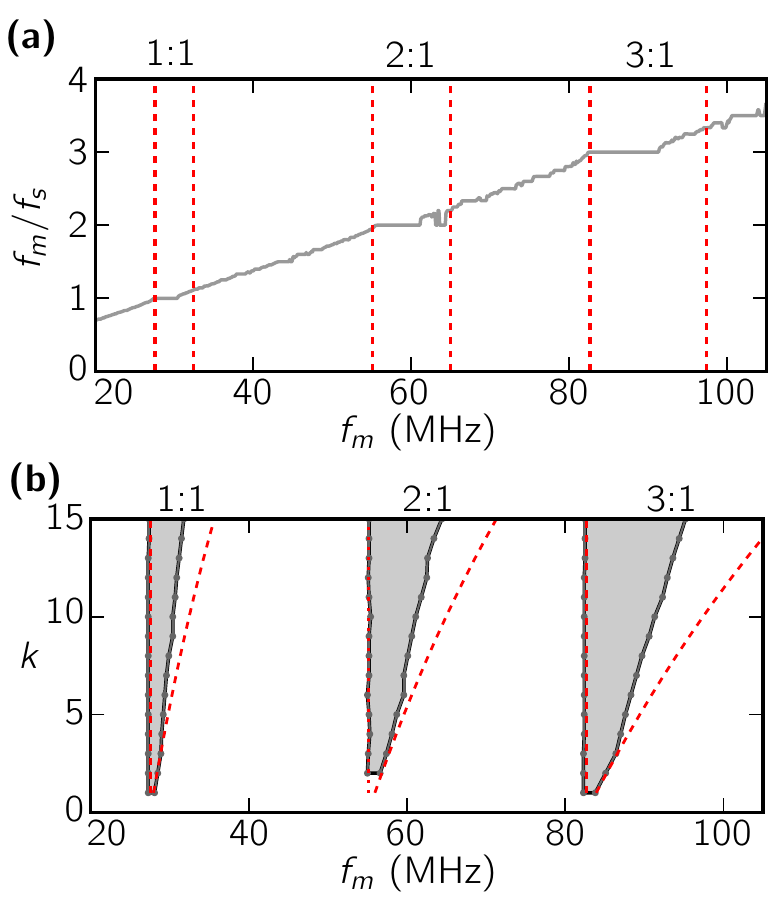}
\caption{
\label{fig:05}
(color online) (a) Devil's stair case and (b) Arnold tongues from numerical simulations, similar to Fig.~\ref{fig:03}. The parameters are the same as in Fig.~\ref{fig:PRE_Fig_4New}. 
}
\end{figure}

The waveforms generated by numerical simulation of the model are shown in Fig.~\ref{fig:04}(b). The control signal $y_{c,ms}$ is obtained by low-pass filtering, similar to Eq.~\eqref{eq:DE_filt}, the Boolean expression $Y_s (t - \tau _{c,ms} )\oplus Y_m (t - \tau _{c,ms} )$ where $Y_{m,s}$ are the Boolean variables associated to continuous variables $y_{m,s}$. The model captures qualitatively well the experimental measurements shown in Fig.~\ref{fig:master_slave}(b). For example, it exhibits similar fast oscillations at the rising edge of output voltage of the slave BPO. With the model, we can also generate the devil's staircase and Arnold tongues as shown in Fig.~\ref{fig:05}(a) and (b). We notice that the simulations and experiments [compare to Fig.~\ref{fig:03}(a) and (b)] agree quantitatively. The widths of the synchronization region $p{:}1$ as a function of $k$ differ only by $14\%$ on average.  

\subsection{C. Synchronization in a bidirectional coupling configuration} 
\begin{figure}[t]
\includegraphics[width=8.5cm]{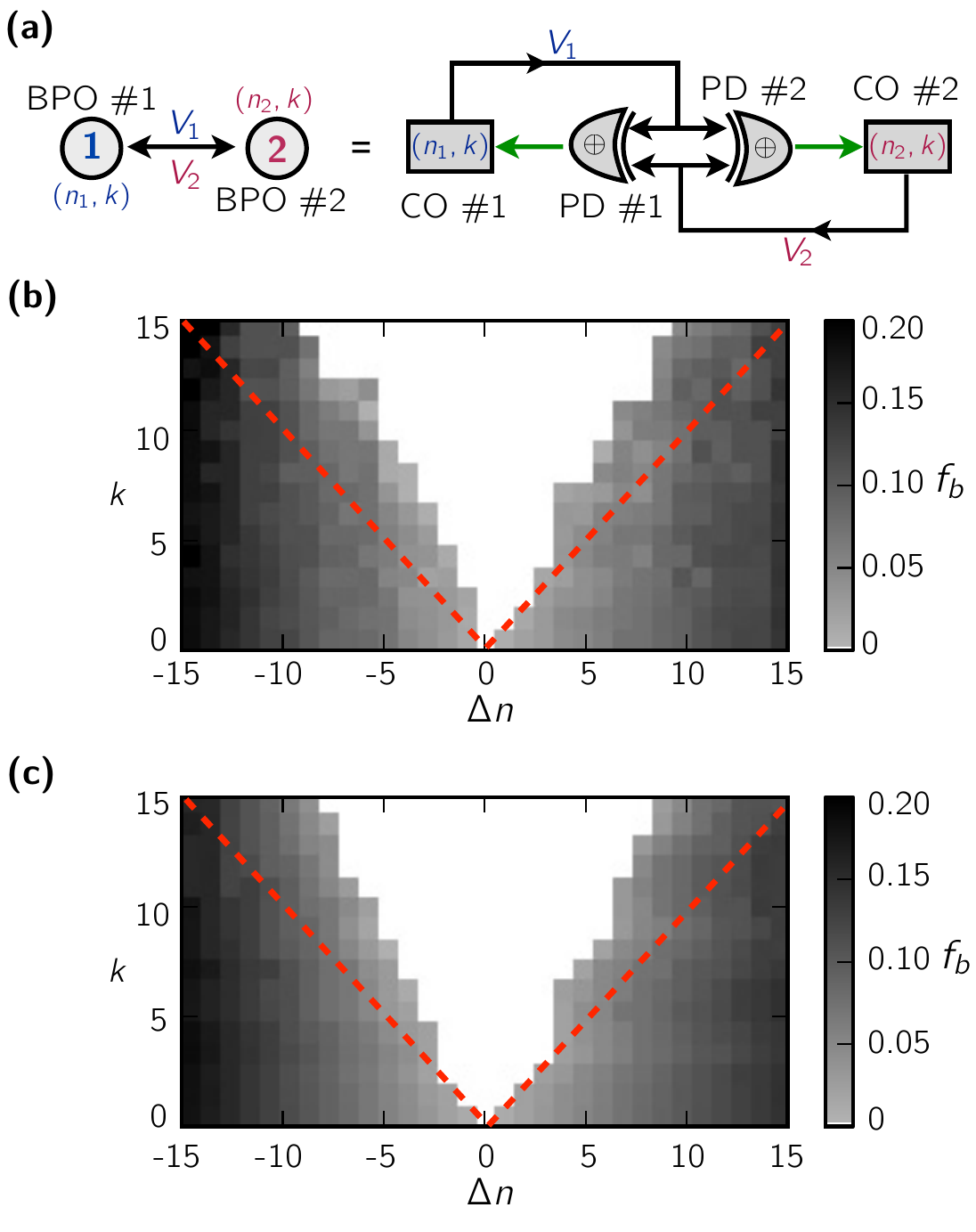}
\caption{(color online) (a) Setup of the of the two mutually coupled oscillators. The coupling is realized with two directed links of equal strength $k$. (b) Experimental and (c) numerical synchronization plane in parameter space $(\Delta n,k)$ measuring the absolute value of the normalized beat frequency $f_b$ for $\Delta n\in \left\{-15,\dots,15\right\}$, $k\in\left\{0,\dots,15\right\}$, and  $n_1=65$ ($n_2=n_1+\Delta n$). The frequency is color-coded; white color indicates normalized beating frequencies $f_b<0.0025$. The red dashed lines indicate the analytic synchronization boundary $k=\left|n_2-n_1\right|$. The parameters are same as in Fig.~\ref{fig:PRE_Fig_4New}.} 
\label{fig:PRE_Fig_9New} 
\end{figure}

To analyze the synchronization properties of BPOs further, we now consider bidirectional coupling of two BPOs. Figure~\ref{fig:PRE_Fig_9New}(a) is a schematic of the experimental setup. The oscillators are coupled symmetrically, \emph{i.e.}, the coupling strength $k$ is identical in both oscillators. The free-running frequency between the oscillators, on the other hand, is detuned by choosing parameters $n_1=65$ and $n_2\in\left\{50,\dots,80\right\}$, where $n_1$ and $n_2$ are proportional to the feedback delay in the two oscillators. A difference of $n_1$ and $n_2$ of $\Delta n=n_2-n_1$ results in detuning of the free-running frequency of the two BPOs of $\Delta f = \Delta n/(2\tau_{LG}n_1n_2)$. Besides varying the frequency detuning, we also change the coupling strength from $k=0$ to $k=15$. 

Following the theoretical framework proposed in the previous section, we model the bidirectional coupling setup by a four-dimensional system of coupled differential equations, according to 
\begin{subequations}
\begin{align}
\tau_{x_{1,2}} \dot x_{1,2}(t) &= -x_{1,2}(t) + [\lnot X_{1,2}(t-\tau_{1,2}(t))],\label{eq:DE_main12}\\
\tau_{y_{1,2}} \dot y_{1,2}(t) &= -y_{1,2}(t) + [\lnot X_{1,2}(t-\tau_{1,2}(t))],\label{eq:DE_filt12}
\end{align}
\end{subequations}
where $(x_{1,2},X_{1,2})$ and $(y_{1,2},Y_{1,2})$ are the pairs of continuous and Boolean variables describing the two coupled BPOs. Parameters $\tau_{x_{1,2}}$, $\tau_{y_{1,2}}$, $\tau_{1,2}(t)$ are analogous to $\tau_{x_s}$, $\tau_{y_s}$, $\tau_s(t)$ in the previous section. 

The two BPOs are considered to be synchronized when the absolute value of the normalized beat frequency $f_b=\left|f_1-f_2\right|/\sqrt{f_1^2+f_2^2}$ of the two BPOs is below $0.25\%$. Coupled oscillators are generally expected to synchronize under large coupling $k$ and small frequency detuning $\Delta f$ \cite{PIK01}. Here, we show that this general property also holds for BPOs. 

We analyze the synchronization properties in parameter space $(\Delta n=n_2-n_1,k)$ of frequency detuning and coupling strength. The experimental and numerical results are shown in Fig.~\ref{fig:PRE_Fig_9New}(b) and (c). In both cases, the synchronization region (white area) is V-shaped; it is maximally extended for small values of $\Delta n$ and large values of coupling strength $k$ as expected for coupled oscillators. Experiment and simulation agree quantitatively. We find that the slope associated with the linear regression for the boundaries of the synchronization region differ approximately by $8\%$. Furthermore, outside of the synchronization region, simulation and experiment differ by less than $5\%$. 

The border of the synchronization region can be approximated with a necessary condition on the coupling strength for synchronization. Specifically, the coupling strength has to exceed the detuning of the two BPOs, \emph{i.e.}, $k\geq\left|n_2-n_1\right|$. The resulting maximal border of synchronization $k=\left|n_2-n_1\right|$ is shown in Fig.~\ref{fig:PRE_Fig_9New}(b) and (c) with dashed lines.  However, this condition is not sufficient to guarantee synchronization, similar to our considerations for uni-directional coupling. Therefore, the triangle delimited by the dashed lines is not entirely filled by the locking region. Interestingly, the analytic synchronization region with bidirectional coupling has a symmetric triangular shape different from the locking region for unidirectional coupling [see Fig.~\ref{fig:03}(b)]. 

\section{IV. Conclusions}
Here, we propose a new experimental paradigm of Boolean phase oscillators with state-dependent delays that is a generalization of previous work on all-digital phase-locked loops. This allows us to study experimentally coupled oscillators using inexpensive digital electronic chips. In our autonomous digital design of the oscillators, we realize a variable coupling strength based on Boolean signals. We observe and analyze rich synchronization phenomena similar to those encountered in coupled phase oscillators, such as asymmetric and symmetric V-shaped synchronization regions in uni- and bidirectional coupling schemes of two coupled oscillators, respectively. Our study also contributes significantly to the experimental investigation of state-dependent delay in a physical system, which is of great current interest \cite{WAL03}.

Recent advances in very-large-scale integration of digital electronics will allow us to coupled Boolean phase oscillators to networks that involve a large number of nodes. As a result, the dynamical system introduced here will play an important role in the future for the study of dynamics of large-scale complex networks experimentally, such as chimera states and dynamical phase transitions.

\section{Acknowledgments}
The authors thank Edward Ott, Michele Girvan, and Eckehard Sch\"oll for useful discussions and gratefully acknowledge the financial support of the U.S. Army Research Office Grants W911NF-12-1-0099 and W911NF-12-1-0424. D.P.R. acknowledges the DFG for financial support in the framework of SFB910.

%
\end{document}